\documentclass[conference]{IEEEtran}
\IEEEoverridecommandlockouts
\usepackage{cite}
\usepackage{amsmath,amssymb,amsfonts}
\usepackage{algorithmic}
\usepackage{graphicx}
\usepackage{textcomp}
\usepackage{xcolor}
\usepackage{caption}
\usepackage{subcaption} % in the preamble
\usepackage{bm}  % In the preamble
\usepackage[utf8]{inputenc}  % For UTF-8 encoding
\usepackage{float}
\usepackage[numbers]{natbib}
\usepackage{acro}
\usepackage[justification=centering]{subcaption}
\renewcommand{\bold}[1]{\bm{#1}}
\usepackage{hyperref} % put this in your preamble
\usepackage{diagbox}
\usepackage{wrapfig}

\def\BibTeX{{\rm B\kern-.05em{\sc i\kern-.025em b}\kern-.08em
    T\kern-.1667em\lower.7ex\hbox{E}\kern-.125emX}}

\DeclareAcronym{SAR}{
	short=SAR,
	long=synthetic aperture radar,
}

\DeclareAcronym{SVA}{
	short=SVA,
	long=spatial variant apodization,
}

\DeclareAcronym{NN}{
	short=NN,
	long=neural network,
}

\DeclareAcronym{RNN}{
	short=RNN,
	long=recurrent neural network,
}

\DeclareAcronym{CNN}{
	short=CNN,
	long=convolutional neural network,
}

\DeclareAcronym{TV}{
	short=TV,
	long=total variation,
}

\DeclareAcronym{PnP}{
	short=PnP,
	long=plug-and-play,
}

\DeclareAcronym{SLC}{
	short=SLC,
	long=single look complex,
}

\DeclareAcronym{PDF}{
	short=PDF,
	long=probability density function,
}

\DeclareAcronym{CAD}{
	short=CAD,
	long=computer aided design,
}

\DeclareAcronym{Sim2Real}{
	short=Sim2Real,
	long=simulation to real,
}

\DeclareAcronym{MAE}{
	short=MAE,
	long=mean absolute error,
}

\DeclareAcronym{PSNR}{
	short=PSNR,
	long=peak signal-to-noise ratio,
}

\DeclareAcronym{SSIM}{
	short=SSIM,
	long=structural similarity index,
}

\DeclareAcronym{FSIM}{
	short=FSIM,
	long=feature similarity index,
}

\DeclareAcronym{LPIPS}{
	short=LPIPS,
	long=learned perceptual image patch similarity,
}

\DeclareAcronym{DPS}{
	short=DPS,
	long=deep posterior sampling,
}

\DeclareAcronym{GT}{
	short=GT,
	long=ground truth,
}

\DeclareAcronym{PL}{
	short=PL,
	long=perceptual loss,
}

\DeclareAcronym{EPL}{
	short=EPL,
	long=edge preserving loss,
}

\DeclareAcronym{ENL}{
	short=ENL,
	long=equivalent number of looks,
}

\DeclareAcronym{SE}{
	short=SE,
	long=Squeeze-and-Excitation,
}
\DeclareAcronym{MLP}{
	short=MLP,
	long=multilayer perceptron,
}

\DeclareAcronym{QPM}{
	short=QPM,
	long=quarter power magnitude,
}

\DeclareAcronym{LUT}{
	short=LUT,
	long=look up table,
}

\begin{document}

\title{Sim2Real SAR Image Restoration: Metadata-Driven Models for Joint Despeckling and Sidelobes Reduction}

% %%%%% add authors and affiliations here %%%%%
% \author[1]{Antoine De Paepe}
% \author[2]{Pascal Nguyen}
% \author[2]{Michael Mabelle}
% \author[1]{Cédric Saleun}
% \author[1]{Antoine Jouadé}
% \author[1]{Jean-Christophe Louvigne}

% \affil[1]{Direction Générale de l'Armement Maîtrise de l'Information, Bruz, France.}
% \affil[2]{Agence Ministérielle pour l'Intelligence Artificielle de Défense, Bruz, France.}

\author{
\IEEEauthorblockN{%
Antoine De Paepe\,\IEEEauthorrefmark{1},
Pascal Nguyen\,\IEEEauthorrefmark{2},
Michael Mabelle\,\IEEEauthorrefmark{2},
C\'edric Saleun\,\IEEEauthorrefmark{1},\\
Antoine Jouad\'e\,\IEEEauthorrefmark{1},
Jean-Christophe Louvigne\,\IEEEauthorrefmark{1}}

\vspace{0.5em}

\IEEEauthorblockA{\IEEEauthorrefmark{1}Direction G\'en\'erale de l'Armement Ma\^\i trise de l'Information, Bruz, France}
\IEEEauthorblockA{\IEEEauthorrefmark{2}Agence Minist\'erielle pour l'Intelligence Artificielle de D\'efense, Bruz, France}
}

\maketitle

\begin{abstract}
\Ac{SAR} provides valuable information about the Earth's surface under all weather and illumination conditions. However, the inherent phenomenon of speckle and the presence of sidelobes around bright targets pose challenges for accurate interpretation of \ac{SAR} imagery. Most existing \ac{SAR} image restoration methods address despeckling and sidelobes reduction as separate tasks. In this paper, we propose a unified framework that jointly performs both tasks using \acp{NN} trained on a realistic \ac{SAR} simulated dataset generated with MOCEM. Inference can then be performed on real \ac{SAR} images, demonstrating effective \ac{Sim2Real} transferability. Additionally, we incorporate acquisition metadata as auxiliary input to the \acp{NN}, demonstrating improved restoration performance.
\end{abstract}


\section*{Notation and Terminology}

Throughout this paper, the term \ac{SAR} image refers broadly to \ac{SAR} data, including both \ac{SLC} images and their amplitude representations. When necessary for clarity, we explicitly distinguish between the two.
For Section~\ref{sec:methodology}, all vectors are represented as column vectors. The symbol ‘$\top$’ denotes the matrix transpose operation. For a given column vector $\bold{s} = [s_1, s_2, \cdots , s_n]^\top \in \mathcal{S}^n$, with $\mathcal{S}$ being either $\mathbb{R}$ or $\mathbb{C}$, $s_k$ denotes the $k$th entry of $\bold{s}$. For the specific case where $\mathcal{S} \triangleq \mathbb{C}$, we write $\bold{s} = \bold{s}^\text{re} + i \bold{s}^\text{im}$, with $\bold{s}^\text{re} \in \mathbb{R}^n$ denoting the real part and $\bold{s}^\text{im} \in \mathbb{R}^n$ denoting the imaginary part. We define $\bold{D}_{\bold{s}} \triangleq \text{diag}(\bold{s}) \in \mathcal{S}^{n\times n}$ as the diagonal matrix whose diagonal entries are given by the components of the vector $\bold{s}$.
\section{Introduction}

\Ac{SAR} imaging has emerged as a crucial remote sensing technology, extensively applied across a range of domains, including environmental monitoring, disaster management, and military surveillance. Its capability to acquire high-resolution imagery under various conditions renders \ac{SAR} a valuable tool for observing and analyzing the Earth's surface.

Despite its many advantages, \ac{SAR} imagery is often compromised by inherent noise—particularly speckle—as well as sidelobes surrounding bright targets, which can obscure critical details. To improve the quality of \ac{SAR} image interpretation, it is essential to address the challenges posed by both speckle noise and the presence of sidelobes.

A variety of sidelobes reduction techniques have been developed, each targeting different aspects of \ac{SAR} signal processing. The most widely used methods are windowing functions \cite{windowsfiltering}, such as Hamming, Hanning, and Kaiser windows, which reduce sidelobes amplitude by smoothing the radar signal during processing, albeit at the cost of some resolution loss. More sophisticated approaches aim to minimize sidelobes levels without significantly compromising resolution or signal strength, such as \ac{SVA} \cite{sva}. However, \ac{SVA} presents several drawbacks: it distorts statistics in speckle-dominated regions, spreads point-like targets across multiple pixels, and introduces a negative bias, making homogeneous areas appear less reflective \cite{abergel2018subpixellic}. Recent advancements in deep learning have introduced \ac{CNN}-based \cite{cnnsidelobes} and bidirectional \ac{RNN}-based \cite{rnnsidelobes} models to mitigate these issues.

Parallel to sidelobes reduction techniques, \ac{SAR} image despeckling has been extensively studied and is now commonly addressed through two main categories of methods: variational approaches and learning-based techniques.

Variational methods typically formulate the problem as the minimization of a cost function balancing a data fidelity term—which accounts for the statistical properties of speckle under the Goodman speckle model \cite{goodman}—and a regularization term, which incorporates prior knowledge of the underlying image \cite{Bioucas2010, Deledalle2017}.

Deep learning-based methods are usually trained on natural image pairs augmented with synthetic speckle noise to effectively learn the despeckling task. SAR-CNN \cite{Chierchia2017} and ID-CNN \cite{Wang2017} were specifically designed to handle the multiplicative nature of speckle noise, inspired by the DnCNN architecture \cite{Zhang2017}. Additionally, UNet architectures \cite{Ronneberger2015}, initially developed for medical image segmentation, have proven highly effective in natural image restoration tasks \cite{Gurrola2021, Jia2021, Fan2022}, and their adaptability has been demonstrated in \ac{SAR} despeckling applications \cite{jiang2018, Ko2022}. These networks excel at capturing multi-scale features, which is crucial for restoring fine details in noisy \ac{SAR} images. Generative adversarial networks (GANs) have also emerged as a powerful tool for \ac{SAR} image restoration \cite{Wang2017idgan, Liu2020}. A more recent development in \ac{SAR} image restoration is the application of conditional diffusion processes \cite{Perera2023, hu2024sar} for despeckling \ac{SAR} images. Finally, self-supervised learning approaches \cite{merlin, Chen2023ASS}, which eliminate the reliance on \ac{GT} data, are emerging as effective solutions for \ac{SAR} image restoration.

Despite the promise of deep learning methods, most are primarily designed to reduce spatially uncorrelated speckle due to their training strategies. However, in real \ac{SAR} images, speckle is often spatially correlated through the \ac{SAR} transfer function. This domain gap can lead to suboptimal restoration in practical applications. Few methods have been developed to address this challenge, such as MuLoG-DRUNet \cite{mendes2024robustness}, a \ac{PnP} method that stands out as the first robust technique to reduce spatially correlated speckle.

In this paper, we propose a novel approach for joint despeckling and sidelobes reduction for \ac{SAR} image restoration using deep learning-based methods. We use MOCEM \cite{mocem} to create a dataset of \ac{SAR} images with access to \ac{GT}, specifically designed to replicate the speckle characteristics and effects of the \ac{SAR} transfer function observed in real-world scenarios. This dataset is used for a supervised learning task, and its consistency with real \ac{SAR} acquisitions enables effective \ac{Sim2Real} \ac{SAR} image restoration. Furthermore, we propose incorporating \ac{SAR} acquisition metadata to enhance the restoration process. By integrating metadata into the deep learning framework, we provide contextual information that allows the \acp{NN} to tailor its restoration strategy according to specific acquisition conditions.

\section{Methodology}\label{sec:methodology}

\subsection{Problem Formulation}

The \ac{SAR} acquisition process involves a sensor, typically mounted on an aircraft or satellite, which acts as an active system that emits electromagnetic waves toward the Earth's surface and captures the reflected signals as a set of complex-valued measurements. These measurements are coherently combined to synthesize a long virtual aperture, forming the \ac{SLC} image $\bold{y} \in \mathcal{Y} \triangleq \mathbb{C}^n$ of the scene reflectivity $\bold{x} \in \mathcal{X} \triangleq \mathbb{R}^n_{+}$ being imaged. For interpretability and practical processing, \ac{SAR} images are also expressed in terms of the amplitude of the \ac{SLC}, defined as $\tilde{\bold{y}}\triangleq|\bold{y}|$, with $\tilde{\bold{y}}\in \tilde{\mathcal{Y}} \triangleq \mathbb{R}^n_{+}$. Owing to the impulse response of the \ac{SAR} system and the coherent nature of radar signals, \ac{SAR} images are often degraded by both sidelobes and spatially correlated speckle. We denote by $\bold{H} \in \mathbb{R}^{n\times n}$ the linear spatial-domain operator associated with the \ac{SAR} transfer function. Under the Goodman model, it is commonly assumed that for a restoration task, the measurement $\bold{y}$, in the absence of electronic noise, follows a circular Gaussian distribution \cite{merlin} with independent entries given by

\begin{equation}\label{eq:sar_model}
(\bold{y}|\bold{H}, \bold{x}) \sim \mathcal{CN}(\bold{0}_{\mathcal{X}},\bold{H}\bold{D}_{\bold{x}}\bold{H}^\top),
\end{equation}

\noindent
where $\mathcal{CN}(\bold{\mu}, \bold{\Sigma})$ denotes a complex circular Gaussian distribution with mean $\bold{\mu}\in\mathbb{R}^n$ and covariance matrix $\bold{\Sigma}\in\mathbb{R}^{n\times n}$. This model can be reparametrized to exhibit the multiplicative nature of speckle as

\begin{equation}\label{eq:sar_mult_model}
\bold{y} = \bold{H}(\mathcal{C}(\bold{x}^{\frac{1}{2}}) \odot \bold{n}), \quad \bold{n} \sim \mathcal{CN}(\bold{0}_{\mathcal{X}}, \bold{I}_{\mathcal{X}}),
\end{equation}

\noindent
where $\mathcal{C} : \mathbb{R}^n \rightarrow \mathbb{C}^{n}$ is the complexification operator such that $\mathcal{C}(\bold{x}) = \bold{x} + i \bold{x}$, and $\bold{n}$ represents the multiplicative speckle noise in its complex form.

When the transfer function $\bold{H}$ is known, image restoration can be achieved by finding the maximum a posteriori (MAP) estimate of $\bold{x}$, i.e.,

\begin{equation}\label{eq:map}
\max_{\bold{x} \in \mathcal{X}} p(\bold{\tilde{y}}|\bold{H}, \bold{x}) \cdot p(\bold{x}),
\end{equation}

\noindent
where the conditional \ac{PDF} $p(\bold{\tilde{y}}|\bold{H}, \bold{x})$ is given by \eqref{eq:sar_model}, and $p(\bold{x})$ is a prior \ac{PDF} on $\bold{x}$, which is generally unknown and replaced (in its post-log form) by a regularizer promoting piecewise smoothness. However, solving the inverse problem in \eqref{eq:map} relies on knowledge of $\bold{H}$, which may be unavailable in many real-world applications—particularly when the processing pipeline or acquisition geometry is proprietary or undisclosed. In such cases, traditional model-based restoration approaches become inapplicable or unreliable due to the absence of an accurate system model, especially in the joint task of despeckling and sidelobes reduction.

To address this limitation, we consider an alternative strategy that leverages a large collection of known \ac{SAR} transfer functions to train \ac{NN} $f_{\bold{\theta}}: \tilde{\mathcal{Y}} \to \mathcal{X}$, parameterized by $\bold{\theta} \in \bold{\Theta}$, which can generalize to unseen or implicitly characterized systems. The training process is formulated as the minimization problem

\begin{equation}\label{eq:sar_training}
\min_{\bold{\theta} \in \bold{\Theta}} \mathbb{E}_{(\bold{x}, \bold{\tilde{y}}) \sim p(\bold{x}, \bold{\tilde{y}})}[\ell(f_{\bold{\theta}}(\bold{\tilde{y}}), \bold{x})],
\end{equation}

\noindent
where $p(\bold{x}, \bold{\tilde{y}}) = \int p(\bold{\tilde{y}}|\bold{H},\bold{x}) p(\bold{H}) p(\bold{x})d\bold{H}$ is the joint \ac{PDF} of $\bold{x}$ and $\bold{\tilde{y}}$, marginalized over $p(\bold{H})$, the \ac{PDF} of possible \ac{SAR} transfer functions, and $\ell : \mathcal{X} \times \mathcal{X} \rightarrow \mathbb{R}$ is a loss function. In this way, the model implicitly learns to invert the composite effect of the transfer function $\bold{H}$ and the speckle noise. This approach enables model-free \ac{SAR} image restoration, particularly suited to operational contexts.

\subsection{Sim2Real: Learning with Simulations for Real-World Restoration}
\begin{figure}[H]
\centering
\begin{subfigure}[b]{0.3\columnwidth}
\includegraphics[width=\linewidth]{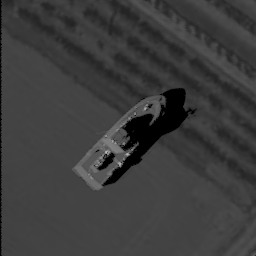}
\centering
\caption{Scene reflectivity $\bold{x}$.}
\end{subfigure}
\hfill
\begin{subfigure}[b]{0.3\columnwidth}
\includegraphics[width=\linewidth]{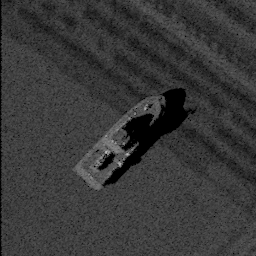}
\centering
\caption{Amplitude of \ac{SLC} $\bold{z}$.}
\end{subfigure}
\hfill
\begin{subfigure}[b]{0.3\columnwidth}
\includegraphics[width=\linewidth]{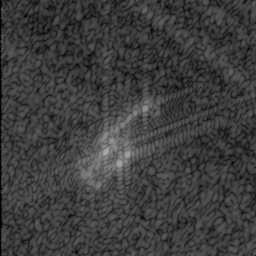}
\centering
\caption{Amplitude of \ac{SLC} $\bold{y}$.}
\end{subfigure}
\caption{Images produced with MOCEM.}
\label{fig:mocem_output}
\end{figure}

In real-world scenarios, obtaining a large dataset of paired samples $(\boldsymbol{x}, \boldsymbol{\tilde{y}}) \sim p(\bold{x}, \bold{\tilde{y}})$ for supervised learning is nearly impossible.
It therefore becomes necessary to overcome this limitation through simulation by generating synthetic training datasets. We propose using simulated \ac{SAR} images obtained with the MOCEM \cite{mocem} software. This simulation tool for modeling \ac{SAR} acquisition enables complete replication of the \ac{SAR} imaging chain, producing high-fidelity \ac{SAR} images from \ac{CAD} models while incorporating fundamental electromagnetic material properties. In particular, it provides access to the observed \ac{SLC} image $\bold{y} \in \mathcal{Y}$, an intermediate \ac{SLC} image $\bold{z} \in \mathcal{Z}$ obtained before the system transfer function $\bold{H}$ is applied (corresponding to $\mathcal{C}(\bold{x}^{\frac{1}{2}}) \odot \bold{n}$ in \eqref{eq:sar_mult_model}), and the scene reflectivity $\bold{x} \in \mathcal{X}$. Although $\bold{x}$ does not represent a physically measurable quantity, it serves as an effective \ac{GT} candidate because it is easily interpretable by human operators. The different images are illustrated in Figure~\ref{fig:mocem_output}.

The MOCEM software also provides, for each scene, detailed metadata $\bold{m} = [m_{1}, \cdots, m_{d}]^\top \in \mathcal{M} \triangleq \mathbb{R}^{d}$, where $d$ is the number of metadata parameters per scene, including radar geometry, squint angle, acquisition resolution, noise level, and other acquisition-specific parameters. This supplementary information can be leveraged to further enhance model performance.

Our approach is to train restoration \acp{NN} exclusively on simulated \ac{SAR} images generated by MOCEM, taking advantage of both the known \ac{GT} reflectivity scenes and the associated metadata. Once trained, these models are directly applied to real \ac{SAR} acquisitions for inference, enabling image restoration despite the absence of paired real data. This \ac{Sim2Real} strategy bridges the gap between synthetic training and operational deployment by exploiting the physical fidelity of simulation to generalize effectively to real-world \ac{SAR} imaging scenarios.

\subsection{Input Variations and Custom Loss Functions for Improved Performance}

For the supervised learning task, the minimization problem \eqref{eq:sar_training} for training the NN $f_{\bold{\theta}}$ can be generalized as

\begin{equation}
\min_{\bold{\theta}}\mathbb{E}_{(\bold{v}, \bold{w}) \sim p(\bold{v}, \bold{w})}[\ell(f_{\bold{\theta}}(\bold{v}), \bold{w})]
\end{equation}

\noindent where $\bold{v} \in \mathcal{V}$ is the input of the NN, $\bold{w} \in \mathcal{W}$ is the target and $p(\bold{v}, \bold{w})$ is the joint distribution. Defining the set of possible inputs, targets, loss functions, and model architectures can have a significant impact on restoration performance. As a model choice, considering the effectiveness of the DRUNet architecture \cite{zhang2021}, originally designed for Gaussian denoising in a \ac{PnP} ADMM framework, we chose to retain it and adapt it for \ac{SAR} image restoration. We also implemented a DRUNet enhanced with \ac{SE} blocks—modules that perform a squeeze operation via global pooling followed by an \ac{MLP} generating channel-wise excitation weights to recalibrate feature maps \cite{hu2019squeezeandexcitationnetworks}—which we call SEDRUNet. We retain $\bold{v} = (\tilde{\bold{y}}, \bold{y}^{\text{re}}, \bold{y}^{\text{im}})$, stacking the amplitude $\tilde{\bold{y}}$ alongside both the real and imaginary parts, $\bold{y}^{\text{re}}$ and $\bold{y}^{\text{im}}$, as the input of the \acp{NN}. Although the amplitude information is contained in the real and imaginary parts of $\bold{y}$, we experimented (results not shown in the paper) and found that using information from these three channels improves performance. The target is set to $\bold{w} = \bold{x}$. We propose training the DRUNet architectures using different loss functions $\ell$: the \ac{MAE}, commonly used in image restoration tasks; the \ac{PL} \cite{johnson2016perceptual}, which compares high-level feature representations extracted from a pretrained network to preserve perceptual quality; and the \ac{EPL} \cite{pandey2018msce}, which emphasizes the retention of sharp edges and fine structures by penalizing differences in image gradients. For the following, we denote by DRUNet \ac{EPL} (resp. DRUNet \ac{PL}) the models trained using the corresponding loss function.

\begin{figure}[H]
\centering
\includegraphics[width=\linewidth]{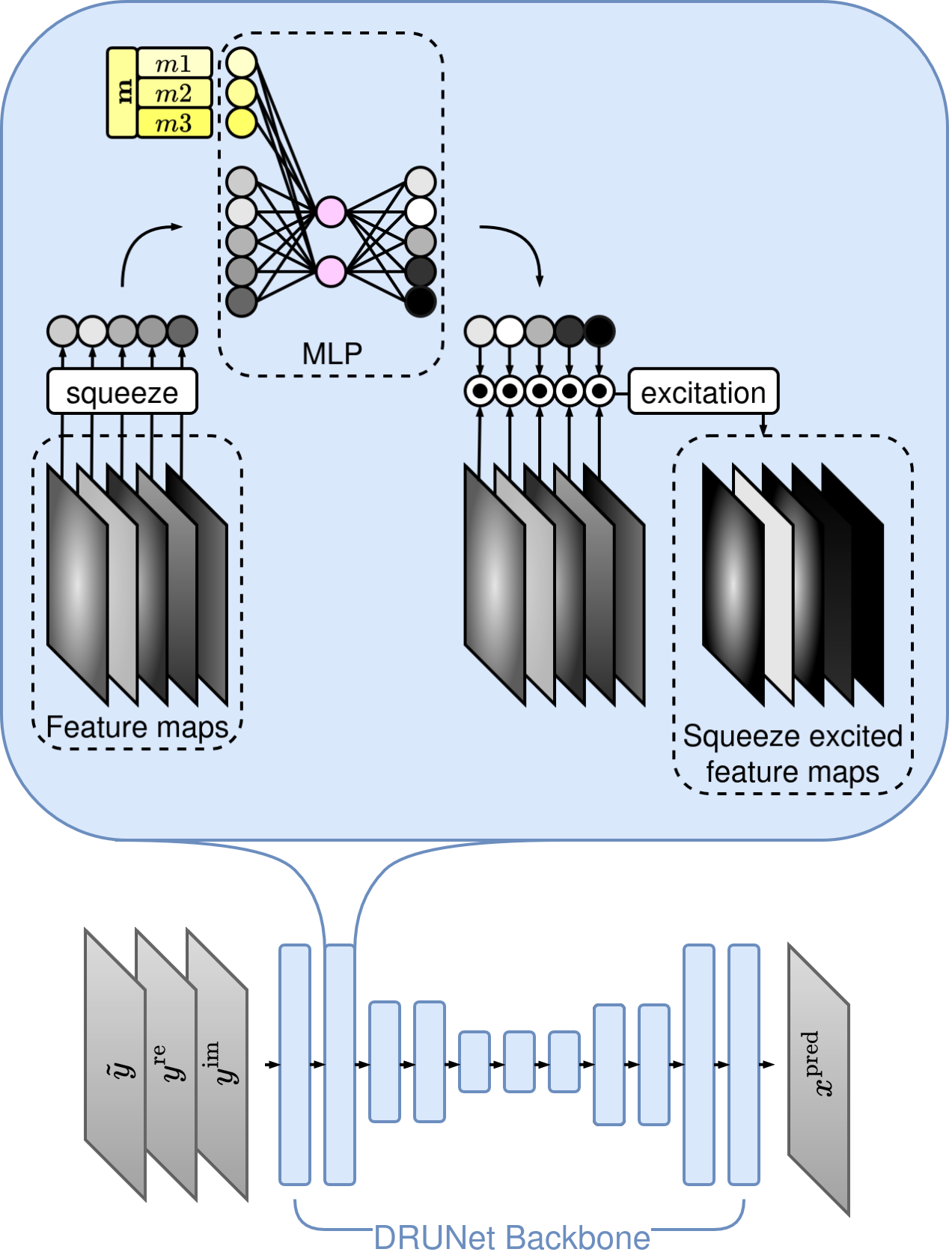}
\caption{Overview of the architecture of M-SEDRUNet.}
\label{fig:metadata_se}
\end{figure}

Finally, we also leveraged the available metadata in the dataset to enhance the model’s restoration. In that context, $\bold{v} = (\tilde{\bold{y}}, \bold{y}^{\text{re}}, \bold{y}^{\text{im}}, \bold{m})$ and $\bold{w} = \bold{x}$ remain the same as before. We propose two different methods to incorporate the metadata. The naive one, which we call M-DRUNet, consists of passing the metadata through input maps, similar to image metadata maps. For each metadata parameter $m_i$, a map of the same spatial dimension as the input amplitude $\tilde{\bold{y}}$ is created and filled with the corresponding single metadata value. The second approach, called M-SEDRUNet, incorporates metadata through \ac{SE} blocks, inspired by the work of Plutenko et al. \cite{plutenko2023metadata}. Metadata are then injected into the \acp{MLP} of the \ac{SE} blocks, enabling it to adapt dynamically to context. Figure \ref{fig:metadata_se} shows the metadata injection principle for M-SEDRUNet.

\section{Experiments}

\subsection{Data Preparation}

\begin{figure*}[ht]
    \centering
    \begin{subfigure}[b]{0.15\textwidth}
        \includegraphics[width=\linewidth]{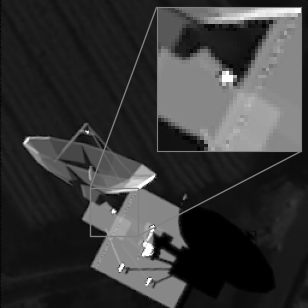}
        \caption{GT}
    \end{subfigure}
    \hfill
    \begin{subfigure}[b]{0.15\textwidth}
        \includegraphics[width=\linewidth]{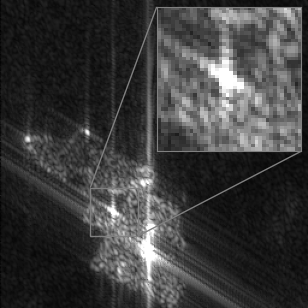}
        \caption{SAR}
    \end{subfigure}
    \hfill
    \begin{subfigure}[b]{0.15\textwidth}
        \includegraphics[width=\linewidth]{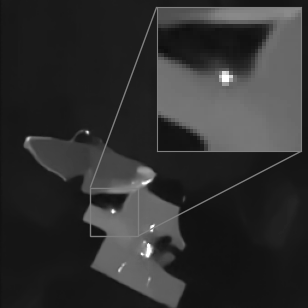}
        \caption{DRUNet}
    \end{subfigure}
    \hfill
    \begin{subfigure}[b]{0.15\textwidth}
        \includegraphics[width=\linewidth]{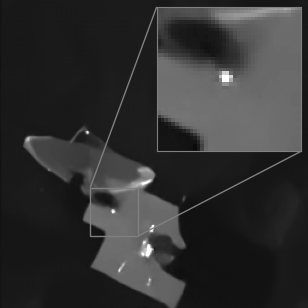}
        \caption{DRUNet EP}
    \end{subfigure}
    \hfill
    \begin{subfigure}[b]{0.15\textwidth}
        \includegraphics[width=\linewidth]{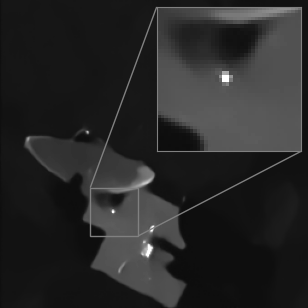}
        \caption{DRUNet PL}
    \end{subfigure}
    \hfill
    \begin{subfigure}[b]{0.15\textwidth}
        \includegraphics[width=\linewidth]{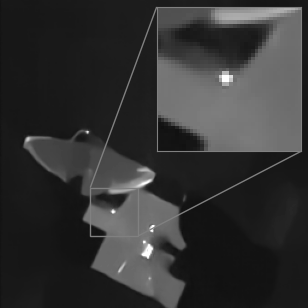}
        \caption{SEDRUNet}
    \end{subfigure}

    \vspace{1ex}

    \begin{subfigure}[b]{0.15\textwidth}
        \includegraphics[width=\linewidth]{figure/transparent.png}
    \end{subfigure}
    \hfill
    \begin{subfigure}[b]{0.15\textwidth}
        \includegraphics[width=\linewidth]{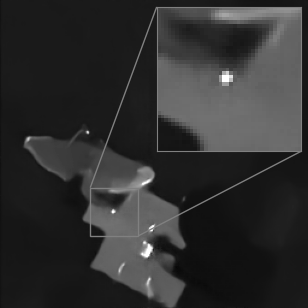}
        \caption{SARCAM}
    \end{subfigure}
    \hfill
    \begin{subfigure}[b]{0.15\textwidth}
        \includegraphics[width=\linewidth]{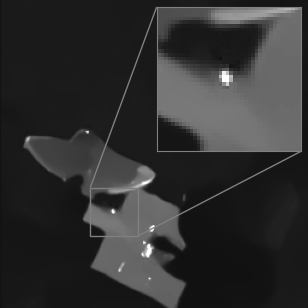}
        \caption{IRNeXt}
    \end{subfigure}
    \hfill
    \begin{subfigure}[b]{0.15\textwidth}
        \includegraphics[width=\linewidth]{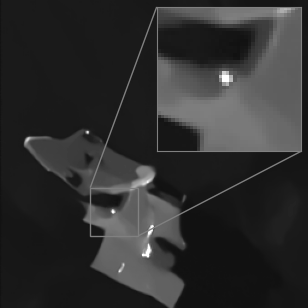}
        \caption{AdaIR}
    \end{subfigure}
    \hfill
    \begin{subfigure}[b]{0.15\textwidth}
        \includegraphics[width=\linewidth]{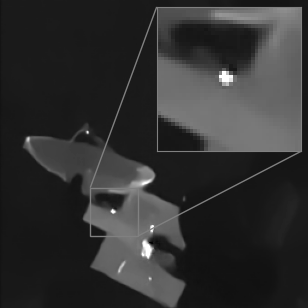}
        \caption{M-DRUNet}
    \end{subfigure}
    \hfill
    \begin{subfigure}[b]{0.15\textwidth}
        \includegraphics[width=\linewidth]{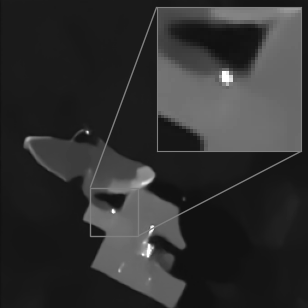}
        \caption{M-SEDRUNet}
    \end{subfigure}

    \caption{\Ac{GT} and and restored simulated \ac{SAR} images.}
    \label{fig:simulated_sar_restoration}
\end{figure*}

As previously mentioned, we used a dataset of \ac{SAR} images simulated using the MOCEM software. The \ac{GT} images of size $n = 256 \times 256$ are composed of background maps into which various \ac{CAD} models are inserted. To ensure scene diversity, a total of 78 distinct \ac{CAD} models—representing both military and civilian objects such as airplanes, boats, antennas, land vehicles, and buildings—are randomly placed with varying scales and orientations onto background images. The background images themselves depict a range of environments, including forests, roads, fields, and airports. From these composite scenes, MOCEM is used to simulate realistic \ac{SAR} acquisitions. It is important to note that, in this case, MOCEM accurately models the electromagnetic scattering behavior only for the \ac{CAD} objects; the background images do not reflect true radar backscattering properties and are included solely to enhance contextual realism. To ensure diversity in acquisition scenarios, the simulation process involves uniform sampling of scene and sensor parameters, including radar acquisition settings such as squint angle, spatial resolution, and noise level. This variability contributes to the generation of a rich and diverse dataset, better preparing the trained \acp{NN} for real-world generalization. The final generated dataset consists of 5,000 4-tuples $(\bold{y}, \bold{z}, \bold{x}, \bold{m}) \in \mathcal{Y} \times \mathcal{Z} \times \mathcal{X} \times \mathcal{M}$, which were split into 4,631 training samples and 369 validation samples. The split was performed by isolating specific acquisition parameter ranges, as shown in Table~\ref{table:mocem_params}, to ensure that the validation set contains conditions not present in the training set. This design explicitly tests the \acp{NN} ability to generalize to unseen radar acquisition configurations. Note that all images are standardized according to the statistics of the training split. The same standardization procedure is applied individually to each metadata variable.

\begin{table}[h]
\renewcommand{\arraystretch}{1.3} % Increase row height
\centering
\begin{tabular}{|c|c|c|}
\hline
\diagbox{Metadata}{Dataset} & Training set & Validation set \\
\hline
Bearing ($^\circ$)& $[0, 150[ \, \cup \, ]155, 360]$ & $[150, 155]$ \\
Incidence ($^\circ$)& $[15, 55[ \, \cup \, ]55.7, 75]$ & $[55, 55.7]$ \\
Squint ($^\circ$) & $[0, 15[ \, \cup \, ]15.5, 45]$ & $[15, 15.5]$ \\
Resolution (m$^2$) & $[0.2, 0.35[ \, \cup \, ]0.36, 0.6]$ & $[0.35, 0.36]$ \\
Noise level (db) & $[-40, -32[ \, \cup \, ]-31.7, -20]$ & $[-32, -31.7]$ \\
\hline
\end{tabular}
\caption{Partitioning of the training and validation sets based on non-overlapping metadata intervals of simulated acquisitions.}\label{table:mocem_params}
\end{table}
\noindent
Finally, we used real UMBRA\footnote{https://umbra.space/open-data/} and CAPELLA\footnote{https://www.capellaspace.com/earth-observation/gallery} \ac{SLC} images as a test set. These images have statistics properties similar to those seen in training, thereby reducing the domain gap.

\subsection{Experimental Settings}

For performance evaluation, we compare our proposed methods specifically with SARCAM \cite{ko2021sar}, a deep learning architecture designed for \ac{SAR} image despeckling, as well as with AdaIR \cite{cui2024adair} and IRNeXt \cite{cui2023irnext}, originally developed for general image restoration. They are included in our comparison as they are well suited to the joint \ac{SAR} despeckling and sidelobes reduction problem. In addition, we consider MERLIN \cite{merlin} and MuLoG-DRUNet \cite{mendes2024robustness}, well-established and effective approaches for real \ac{SAR} data despeckling. Their inclusion allows us to benchmark our method against robust despeckling baselines, even though they do not explicitly target sidelobes reduction. Note that both models are evaluated only on real \ac{SAR} images using publicly available pretrained weights.

All models are trained using the Adam optimizer with an initial learning rate of $1 \times 10^{-4}$, coupled with an exponential decay schedule. Training is conducted for 200 to 400 epochs, depending on when convergence is observed on the validation set. The batch size is adjusted according to available GPU memory. To improve generalization, data augmentation techniques such as random flips and rotations are applied. We use classical image restoration metrics such as \ac{PSNR} and \ac{SSIM} to evaluate image quality. In addition, we include \ac{FSIM} \cite{zhang2011fsim} and \ac{LPIPS} \cite{zhang2018unreasonable}, which capture perceptual fidelity and better align with human visual perception. To evaluate despeckling performance on real \ac{SAR} images, we use the \ac{ENL}, which measures the level of speckle noise reduction in homogeneous \ac{SAR} image regions. No quantitative metric for sidelobes reduction is used, as no established metric currently exists.

\subsection{Results on Simulated \ac{SAR} Images}

\begin{table}[h]
\renewcommand{\arraystretch}{1.2} % Increase row height
\centering
\begin{tabular}{|l|c|c|c|c|}
\hline
\diagbox{Model}{Metric}  & PSNR $\uparrow$ & SSIM $\uparrow$ & FSIM $\uparrow$ & LPIPS $\downarrow$ \\
\hline
SARCAM & 29.915 & 0.897 & 0.877 & 0.347 \\
IRNeXt & 30.023 & 0.899 & 0.874 & 0.342 \\
AdaIR & 30.040 & 0.899 & 0.878 & \textbf{0.333} \\
DRUNet & 29.794 & 0.897 & 0.876 & 0.338 \\
SEDRUNet & 30.012 & 0.899 & 0.874 & 0.339 \\
DRUNet \ac{PL} & 29.919 & 0.897 & 0.879 & 0.341 \\
DRUNet \ac{EPL} & 29.836 & 0.896 & \textbf{0.880} & 0.340 \\
M-DRUNet & 30.272 & 0.899 & 0.878 & 0.339 \\
M-SEDRUNet & \textbf{30.461} & \textbf{0.901} & 0.874 & 0.335 \\
\hline
\end{tabular}
\caption{Quantitative evaluation on the simulated SAR validation set.}
\end{table}

In terms of \ac{PSNR}, the best-performing models are those incorporating metadata, which highlights the impact of metadata integration on the image restoration task. On the other hand, according to perceptual metrics such as \ac{SSIM}, \ac{FSIM}, and \ac{LPIPS}, all models demonstrate comparable performance. The comparison between DRUNet and SEDRUNet and their metadata-driven counterparts demonstrates that metadata injection enhances model performance, even though the improvement margin remains modest. From a visual standpoint, the restored images appear similar across all models (see Figure~\ref{fig:simulated_sar_restoration}). Speckle noise is completely suppressed, and sidelobes are clearly eliminated in all restored outputs.

\subsection{Results on Real SAR Images}

\begin{wraptable}{r}{0.2\textwidth} % 'r' for right side, width of table box
\footnotesize
\renewcommand{\arraystretch}{1.2} % Increase row height
\centering
\begin{tabular}{|l|c|}
\hline
\diagbox{Model}{Metric}  & ENL $\uparrow$ \\
\hline
MERLIN  & 257.6\\
MuLoG DRUNet & \textbf{367.7}\\
SARCAM & 121.9 \\
IRNeXt & 242.1 \\
AdaIR & 173.1 \\
DRUNet & 83.35\\
SEDRUNet & 175.7\\
DRUNet \ac{PL} & 73.71\\
DRUNet \ac{EPL} & 60.15\\
M-DRUNet & 121.2\\
M-SEDRUNet & 286.2\\
\hline
\end{tabular}
\caption{Quantitative evaluation on real SAR images.}
\label{tab:enl_comparison}
\end{wraptable}

Table \ref{tab:enl_comparison} summarizes the quantitative evaluation of all models on real \ac{SAR} imagery using the \ac{ENL} metric. The despeckling-oriented methods MERLIN and MuLoG-DRUNet exhibit strong performance, with MuLoG-DRUNet achieving the best overall \ac{ENL}, confirming their effectiveness on real data. The metadata-driven M-SEDRUNet attains the second-highest \ac{ENL}, surpassing all DRUNet-based baselines (DRUNet, DRUNet \ac{EPL}, DRUNet \ac{PL}) and demonstrating the benefit of incorporating acquisition metadata for improved generalization to real imagery. IRNeXt and AdaIR also yield competitive results, indicating that general-purpose restoration models can adapt well to \ac{SAR} despeckling. Overall, these results highlight that metadata-aware architectures can approach the performance of specialized despeckling methods while simultaneously addressing sidelobes reduction.

While \ac{ENL} does not capture other critical aspects of \ac{SAR} image restoration, such as the suppression of sidelobes or detail preservation in structured areas, relying solely on \ac{ENL} may provide an incomplete assessment of a model’s performance. Visual inspection therefore remains crucial for evaluating the overall quality of \ac{SAR} image restoration.

\begin{figure*}[ht]
    \centering
    \begin{subfigure}[b]{0.16\textwidth}
        \includegraphics[width=\linewidth]{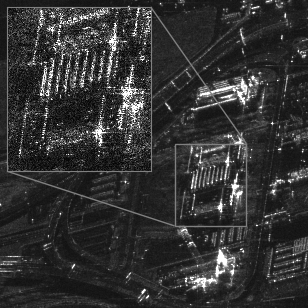}
        \caption{SAR}
    \end{subfigure}
    \hfill
    \begin{subfigure}[b]{0.16\textwidth}
        \includegraphics[width=\linewidth]{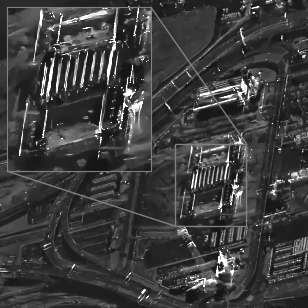}
        \caption{DRUNet}
    \end{subfigure}
    \hfill
    \begin{subfigure}[b]{0.16\textwidth}
        \includegraphics[width=\linewidth]{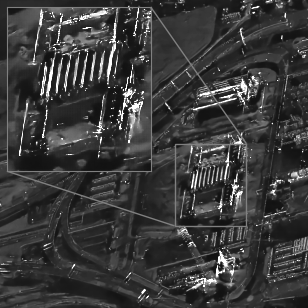}
        \caption{DRUNet EPL}
    \end{subfigure}
    \hfill
    \begin{subfigure}[b]{0.16\textwidth}
        \includegraphics[width=\linewidth]{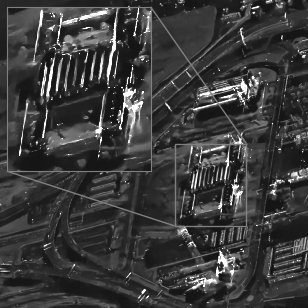}
        \caption{DRUNet PL}
    \end{subfigure}
    \hfill
    \begin{subfigure}[b]{0.16\textwidth}
        \includegraphics[width=\linewidth]{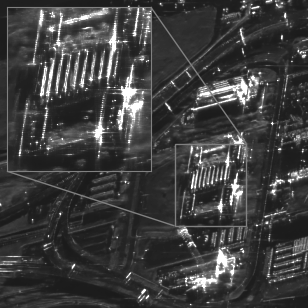}
        \caption{MERLIN}
    \end{subfigure}
    \hfill
    \begin{subfigure}[b]{0.16\textwidth}
        \includegraphics[width=\linewidth]{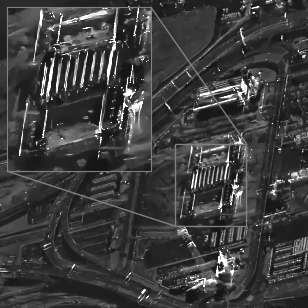}
        \caption{SEDRUNet}
    \end{subfigure}

    \vspace{1ex}

    \begin{subfigure}[b]{0.16\textwidth}
        \includegraphics[width=\linewidth]{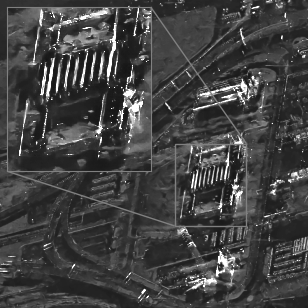}
        \caption{SARCAM}
    \end{subfigure}
    \hfill
    \begin{subfigure}[b]{0.16\textwidth}
        \includegraphics[width=\linewidth]{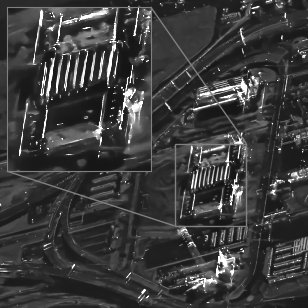}
        \caption{IRNeXt}
    \end{subfigure}
    \hfill
    \begin{subfigure}[b]{0.16\textwidth}
        \includegraphics[width=\linewidth]{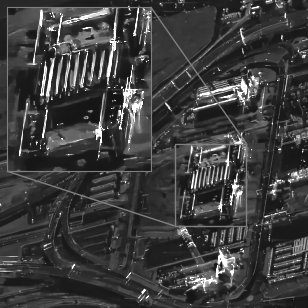}
        \caption{AdaIR}
    \end{subfigure}
    \hfill
    \begin{subfigure}[b]{0.16\textwidth}
        \includegraphics[width=\linewidth]{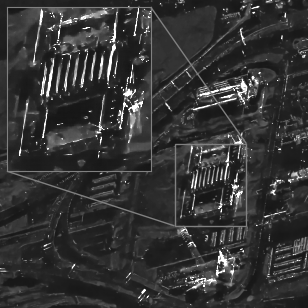}
        \caption{M-DRUNet}
    \end{subfigure}
    \hfill
    \begin{subfigure}[b]{0.16\textwidth}
        \includegraphics[width=\linewidth]{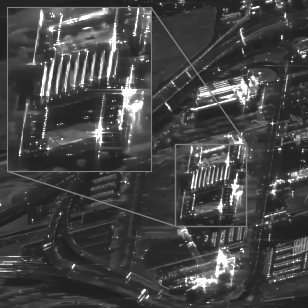}
        \caption{MuLoG DRUNet}
    \end{subfigure}
    \hfill
    \begin{subfigure}[b]{0.16\textwidth}
        \includegraphics[width=\linewidth]{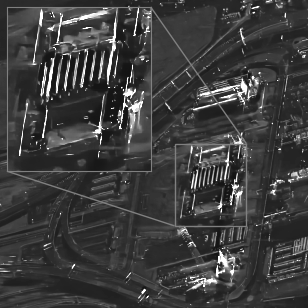}
        \caption{M-SEDRUNet}
    \end{subfigure}

    \caption{Restored images from real UMBRA SLC.}
    \label{fig:real_sar_restoration}
\end{figure*}

\begin{figure}[H]
    \centering
    \begin{subfigure}[b]{0.3\columnwidth}
        \includegraphics[width=\linewidth]{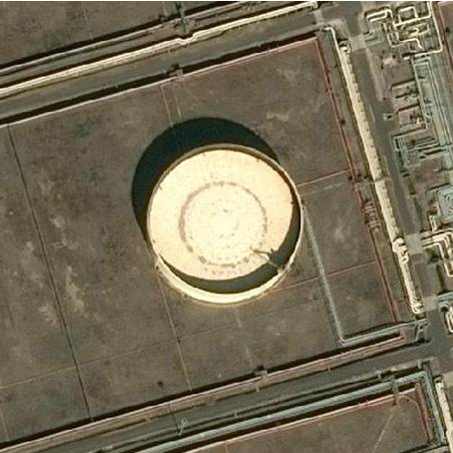}
        \caption{}
    \end{subfigure}
    \hfill
    \begin{subfigure}[b]{0.3\columnwidth}
        \includegraphics[width=\linewidth]{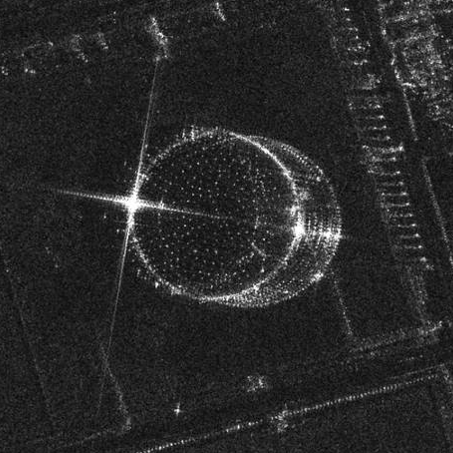}
        \caption{}
    \end{subfigure}
    \hfill
    \begin{subfigure}[b]{0.3\columnwidth}
        \includegraphics[width=\linewidth]{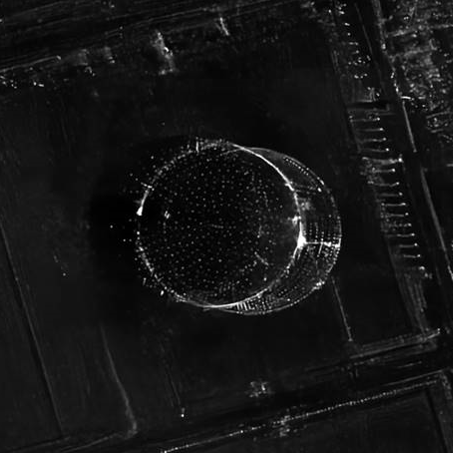}
        \caption{}
    \end{subfigure}
    \caption{Optical GT (a) alongside CAPELLA SAR image (b) and restored image with SEDRUNet (c).}
    \label{fig:capella}
\end{figure}

Regarding the results shown in Figure~\ref{fig:real_sar_restoration}, the despeckling-only methods MERLIN and MuLoG-DRUNet exhibit strong speckle suppression performance on real \ac{SAR} images but fail to remove sidelobe artefacts, as they are not designed for joint restoration. In contrast, the proposed approaches, along with general-purpose models such as IRNeXt and AdaIR, demonstrate both despeckling and sidelobes reduction capabilities. While most methods effectively smooth homogeneous regions, IRNeXt tends to over-smooth certain areas, achieving strong sidelobes suppression at the cost of texture fidelity. DRUNet-based models generally preserve more structural details, though some loss is observed in regions with repetitive patterns, particularly for DRUNet \ac{PL}. The metadata-enhanced variants maintain restoration quality comparable to their non-metadata counterparts, offering a balanced trade-off between noise suppression and detail preservation. Additional visual results are presented in Figure~\ref{fig:capella}.

\begin{figure}[H]
    \centering
    \begin{subfigure}[b]{0.3\columnwidth}
        \includegraphics[width=\linewidth]{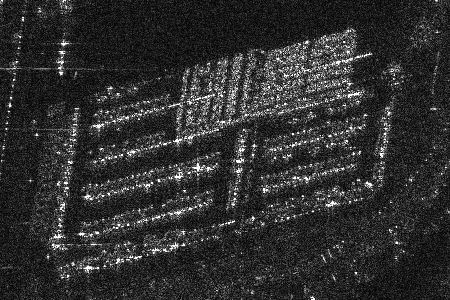}
        \caption{}
    \end{subfigure}
    \hfill
    \begin{subfigure}[b]{0.3\columnwidth}
        \includegraphics[width=\linewidth]{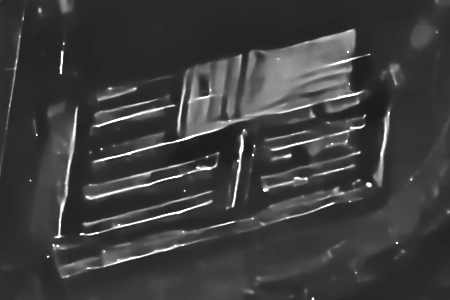}
        \caption{}
    \end{subfigure}
    \hfill
    \begin{subfigure}[b]{0.3\columnwidth}
        \includegraphics[width=\linewidth]{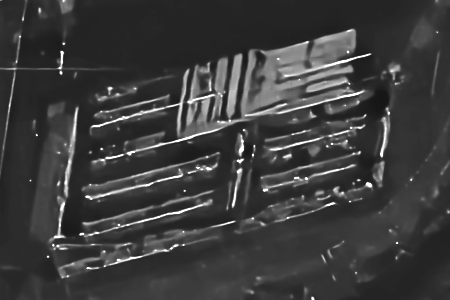}
        \caption{}
    \end{subfigure}
    \caption{UMBRA SAR image (a) and restorations with M-SEDRUNet using the real sensor resolution set to 0.40m$^2$ (b) and a different one set to 0.25m$^2$ (c).}
    \label{fig:metadata_effects}
\end{figure}

Even if the effects of metadata are not always directly observable in the restored images, an interesting outcome is that metadata-driven models can implicitly interpret the underlying physics of the acquisition parameters for the restoration task. For example, injecting a resolution parameter that differs from the one used by the sensor to acquire the original \ac{SAR} image can influence the restored image’s effective resolution. Figure~\ref{fig:metadata_effects} clearly illustrates that providing a higher resolution parameter leads to visibly enhanced image detail. This capability, while promising, has some drawbacks. In particular, it can introduce new sidelobes or artefacts when the injected metadata significantly deviate from the original acquisition parameters. Nonetheless, this adaptive functionality opens new possibilities for dynamic image analysis and restoration tuning. Similar behaviors are observed when varying the incidence angle.

\section{Discussion}\label{sec:discussion}

This study demonstrates the potential of the \ac{Sim2Real} approach, as well as the integration of \ac{SAR} acquisition parameters into the \acp{NN} to enhance performance. However, several limitations must still be addressed.

The injection of metadata into DRUNet and SEDRUNet has shown improved performance, particularly in simulated environments where \ac{GT} data are accessible. Comparisons using identical architectural baselines confirm the benefits of this approach. However, we have not yet explored metadata injection in other advanced models such as AdaIR and IRNeXt. Integrating metadata into these architectures could potentially further enhance their performance and remains an avenue for future investigation.

While the current results confirm that metadata can guide the restoration process by modulating the network’s behavior according to acquisition conditions, the quantitative characterization of this control remains limited. Specifically, the trade-off between the degree of metadata influence and the risk of artefact introduction has not yet been systematically evaluated. Future work should therefore investigate how varying the weighting or conditioning strength of metadata impacts quantitative metrics and perceptual fidelity.

One major challenge is the domain gap between the simulated images used during training and real \ac{SAR} images. For certain \ac{SAR} image providers, such as TerraSAR-X\footnote{\url{https://earth.esa.int/eogateway/missions/terrasar-x-and-tandem-x/sample-data}}, the statistical properties of real images deviate from those seen in training, leading to poor image restoration results. Currently, no effective solution has been identified to reconcile these statistical discrepancies, highlighting the limitations of supervised training strategies in this context. This motivates the exploration of unsupervised approaches. Recently, \ac{DPS} approaches \cite{chung2022diffusion} have gained attention for solving inverse problems, where prior information is learned using diffusion models. Methods such as BlindDPS \cite{chung2023parallel} aim to jointly estimate the forward operator and the latent clean image $\bold{x}$. Applying such techniques in the \ac{SAR} domain may enable the joint estimation of both the \ac{SAR} transfer function and the underlying image, offering a promising direction for future work.

\section{Conclusion}

In this work, we presented a unified deep learning framework for simultaneous despeckling and sidelobes reduction in \ac{SAR} images. Leveraging the realistic \ac{SAR} simulated dataset generated by MOCEM, which includes access to \ac{GT}, we enabled supervised training that effectively addresses both inherent noise and artefacts in \ac{SAR} imagery. Our results demonstrate that incorporating acquisition metadata as auxiliary input improves restoration performance and enables the guided control of image restoration.

\section*{Acknowledgments}

This study has been carried out with financial support
from the French Direction Générale de l’Armement and
the French Agence Ministérielle pour l'Intelligence Artificielle de Défense.

\bibliographystyle{unsrt}
\bibliography{refs}

\end{document}